\documentclass[pra,twocolumn,superscriptaddress]{revtex4}

\usepackage{amsmath}
\usepackage{latexsym}
\usepackage{amssymb}
\usepackage{pdfpages}
\usepackage{graphicx}
\usepackage{subfigure}
\usepackage[colorlinks=true, citecolor=blue, urlcolor=blue]{hyperref}
\usepackage{float}
\usepackage{epstopdf}
\usepackage{pgffor}
\usepackage{amsfonts}
\usepackage{textcomp}
\usepackage{mathpazo}
\usepackage{comment}
\usepackage{braket}
\usepackage{bbm}
\usepackage{xcolor}
\definecolor{myurlcolor}{rgb}{0,0,0.8}
\definecolor{mycitecolor}{rgb}{0,0,0.8}
\definecolor{myrefcolor}{rgb}{0,0,0.8}
\usepackage{hyperref}
\usepackage[capitalise]{cleveref}
\hypersetup{colorlinks,
linkcolor=myrefcolor,
citecolor=mycitecolor,
urlcolor=myurlcolor}

\usepackage[draft]{fixme}
\usepackage{amsmath,bbm}
\usepackage{graphicx}
\usepackage{amsfonts}
\usepackage{amssymb}
\usepackage{amsmath, amssymb, amsthm,verbatim,graphicx,bbm}
\usepackage{mathrsfs}
\usepackage{color,xcolor,longtable}

\makeatletter
\usepackage{pifont}
\makeatother

\usepackage{epstopdf}
\usepackage{subfigure}
\usepackage{appendix}

\makeatletter
\usepackage{pifont}
\makeatother

\usepackage{dcolumn}
\usepackage{epstopdf}

\newcommand{\idol}{\ensuremath{\mathbbm 1}}

\newcommand{\tr}{{\rm Tr}}

\begin{document}
\title{Quantum error pre-compensation for quantum noisy channels}
\author{Chengjie Zhang}
\email{chengjie.zhang@gmail.com}
\affiliation{School of Physical Science and Technology, Ningbo University, Ningbo, 315211, China}

\author{Liangsheng Li}
\affiliation{Science and Technology on Electromagnetic Scattering Laboratory, Beijing 100854, China}
\author{Guodong Lu}
\affiliation{School of Physical Science and Technology, Ningbo University, Ningbo, 315211, China}

\author{Haidong Yuan}
\email{hdyuan@mae.cuhk.edu.hk}
\affiliation{Department of Mechanical and Automation Engineering, The Chinese University of Hong Kong, Hong Kong}
\author{Runyao Duan}
\email{duanrunyao@baidu.com}
\affiliation{Institute for Quantum Computing, Baidu Research, Beijing 100193, China}

\begin{abstract}
Most previous efforts of quantum error correction focused on either extending classical error correction schemes to the quantum regime by performing a perfect correction on a subset of errors, or seeking a recovery operation to maximize the fidelity between a input state and its corresponding output state of a noisy channel. There are few results concerning quantum error pre-compensation. Here we design an error pre-compensated input state for an arbitrary quantum noisy channel and a given target output state. By following a procedure, the required input state, if it exists, can be analytically obtained in single-partite systems. Furthermore, we also present semidefinite programs to numerically obtain the error pre-compensated input states with maximal fidelities between the target state and the output state. The numerical results coincide with the analytical results.
\end{abstract}
\date{\today}

\maketitle

\section{INTRODUCTION}
Quantum error correction (QEC) schemes are extremely important for physical quantum information processing systems \cite{Rev1,Rev2,Rev3}, because without suitable error correcting procedures many quantum information protocols cannot be realizable. Therefore, in order to protect quantum information against noise, the basic theory of QEC was developed \cite{Rev4,Rev5,Rev6}, after the seminal papers of Shor \cite{QECC1} and Steane \cite{QECC2}.
In analogy to classical coding for noisy channels, the earliest efforts in QEC have generalized encoding techniques from classical error-correction schemes, and a theory of quantum error correcting codes (QECCs) has been developed \cite{Rev1,Rev2,Rev3,Rev4,Rev5,Rev6,QECC1,QECC2,QECC3,QECC4,QECC5,QECC6,QECC7,QECC8,QECC9,QECC10}. If the noise is not too severe, the input quantum information, which is embedded in a coded subspace, can be exposed to the ravages of a noisy environment and recovered via a designed operation to perfectly correct a set of errors. 

Furthermore, the design of QEC can also be cast as an optimization problem \cite{QER1,QER2,QER3,QER4,QER5,QER6,QER7,QER8,QER9,QER10}. Unlike the QECCs designed for perfect correction,  the quantum error recovery  (QER) methods, as explained in \cite{QER3}, focus on seeking a recovery operation to maximize the fidelity between a input state and its corresponding output state of a noisy channel. Consider a noisy quantum channel $\mathcal{E}$, the goal of any QER scheme is to design a recovery operation $\mathcal{R}$, which maximizes the fidelity between an input state $\varrho$  and its output state $\mathcal{R}[\mathcal{E}(\varrho)]$ \cite{QER3}. This optimization problem can be solved by a semidefinite program (SDP) \cite{SDP}. 

\begin{figure}
\includegraphics[scale=0.73]{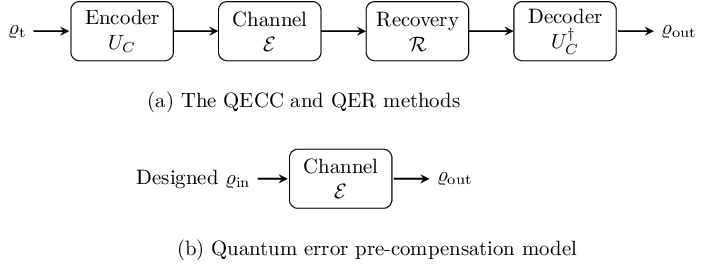}
\caption{Comparison of (a) the QECC and QER methods \cite{QER3} (usually $\varrho_{\rm in}=\varrho_{\rm t}$), with (b) the proposed quantum error pre-compensation model (usually $\varrho_{\rm in}\neq\varrho_{\rm t}$).}\label{fig1}
\end{figure}

The QECC and QER methods are designed to perform recovery operations \textit{after} errors happened. Is there any method \textit{before} errors happened? Actually, Ref. \cite{Knill} has introduced active methods for protecting quantum information against errors, in which they have proposed to use a quantum operation \textit{before} errors happened. Subsequently, the active protecting methods have been formalized in Ref. \cite{Knill2}, and further developed in Ref. \cite{Viola}.

However, the methods \textit{before} errors happened are much less than the methods \textit{after} errors happened. We will propose quantum error pre-compensation (QEPC) scheme which is another method \textit{before} errors happened. In Fig.~\ref{fig1}, we compare the QECC and QER methods with the QEPC model. In the QECC and QER methods, if Alice (sender) would like to send a target state $\varrho_{\rm t}$ to Bob (receiver) via a quantum noisy channel $\mathcal{E}$, she will use $\varrho_{\rm t}$ as the input state, i.e., $\varrho_{\mathrm{in}}=\varrho_{\rm t}$ \cite{QER3}. However, in the QEPC model, we design an error pre-compensated input state $\varrho_{\mathrm{in}}$, such that $\varrho_{\rm out}:=\mathcal{E}(\varrho_{\mathrm{in}})=\varrho_{\rm t}$ or the output state $\varrho_{\rm out}$ is as close as possible to the target output state $\varrho_{\rm t}$. The input state $\varrho_{\mathrm{in}}$, in general, is not equal to the  target state $\varrho_{\rm t}$, i.e., $\varrho_{\mathrm{in}}\neq\varrho_{\rm t}$. The QEPC model is error suppression rather than error correction procedure. One of the motivations of QEPC model is that it would be useful in quantum communications with photonic qubits, such as quantum key distribution via optical fibers. Since large multi-photon entangled states are hard to realize in experiments, previous methods which will use large multi-photon entangled states, like QECC or decoherence-free subspace methods, may not work well, but the QEPC method becomes feasible.

Here we design an error pre-compensated input state $\varrho_{\mathrm{in}}$ for an arbitrary fixed quantum noisy channel $\mathcal{E}$ with a given target output state $\varrho_{\rm t}$. If the required input state $\varrho_{\mathrm{in}}$ exists, it can be analytically obtained by following the procedure in Fig.~\ref{fig2}. Furthermore, we also present two semidefinite programs to numerically obtain the error pre-compensated input states.  The numerical results coincide with the analytical results. If the required input state $\varrho_{\mathrm{in}}$ does not exist, one can use the second semidefinite program to numerically obtain the best input state $\varrho_{\mathrm{in}}$, which maximizes the fidelity between the target state $\varrho_{\rm t}$ and the output state $\mathcal{E}(\varrho_{\mathrm{in}})$.

\section{Analytically design error pre-compensated input states for quantum channels}
Suppose that there is a quantum channel between Alice and Bob. The quantum channel can be viewed as a completely positive trace preserving (CPTP) map $\mathcal{E}$,  with the output state corresponding to an input state $\varrho$ being written in a Kraus form \cite{Rev1},
\begin{equation}\label{CP}
\mathcal{E}(\varrho)=\sum_i K_i\varrho K_i^\dag,
\end{equation}
where $K_i$ are operators satisfying the completeness relation $\sum_i K_i^\dag K_i=\idol$.

\subsection{Single-partite systems}
It is worth noticing that the complete information of this CPTP map $\mathcal{E}$ can be measured by quantum process tomography \cite{Rev1,qpt1,qpt2}, and thus Alice and Bob can obtain full information of $\{K_i\}$ (we assume that once the quantum channel has been set up, it is fixed). If Alice would like to send a special target state $\varrho_{\rm t}$ to Bob via a given quantum channel $\mathcal{E}$, she must design an input state $\varrho_{\mathrm{in}}$ for error pre-compensation such that $\varrho_{\rm t}=\mathcal{E}(\varrho_{\mathrm{in}})$. Generally, the designed input state $\varrho_{\mathrm{in}}$ is different from the target state $\varrho_{\rm t}$, since the quantum channel $\mathcal{E}$ between Alice and Bob is probably a noisy channel. The input state $\varrho_{\mathrm{in}}$, however, may not exist. If $\varrho_{\mathrm{in}}$ exists, it may not be unique. We will discuss all the cases which depend on $\mathcal{E}$ and the target state $\varrho_{\rm t}$.

Hereafter, we will use the notation $\Ket{A}$ as \cite{horn,D'Ariano}
\begin{equation}\label{}
  \Ket{A}:=A\otimes\idol\sum_i|ii\rangle=\sum_{ij}A_{ij}|ij\rangle,
\end{equation}
with $\sum_i|ii\rangle$ being the unnormalized maximally entangled state between subsystems A and B,  the operator $A=\sum_{ij}A_{ij}|i\rangle\langle j|$ which relates the vector $\Ket{A}$ and the operator $A$. Now we focus on our main question:

\textit{Suppose that Alice and Bob share a quantum channel $\mathcal{E}$, described by Eq. (\ref{CP}), and Alice and Bob obtain all the information of this quantum channel in advance. If Alice would like to send a special target state $\varrho_{\rm t}$ to Bob, what input state should Alice choose?}

To answer the above question, we assume that there exists an input state $\varrho_{\mathrm{in}}$ such that
\begin{equation}\label{one-party}
    \varrho_{\rm t}=\mathcal{E}(\varrho_{\mathrm{in}})=\sum_i K_i \varrho_{\mathrm{in}} K_i^{\dag},
\end{equation}
which is equivalent to \cite{horn,D'Ariano}
\begin{equation}\label{vec}
    \Ket{\varrho_{\rm t}}=\Ket{\sum_i K_i \varrho_{\mathrm{in}} K_i^{\dag}}=\sum_i K_i\otimes K_i^*\Ket{\varrho_{\mathrm{in}}},
\end{equation}
the last equation holds due to the definition of $\Ket{A}$, with detailed proof shown in the Appendix A. Therefore, there are several cases for the choice of Alice's input state depending on the target state $\varrho_{\rm t}$ and the matrix $M:=\sum_i K_i\otimes K_i^*$.

\textit{Case (1).} The matrix $M:=\sum_i K_i\otimes K_i^*$ has an inverse matrix $M^{-1}$  (i.e., its determinant $\det M \neq0$). Since $M^{-1}$ exists, from Eq. (\ref{vec}) we have
\begin{equation}\label{}
    \Ket{\varrho_{\mathrm{in}}}=M^{-1} \Ket{\varrho_{\rm t}},
\end{equation}
and from $\Ket{\varrho_{\mathrm{in}}}$ one can obtain $\varrho_{\mathrm{in}}$ by using $A=\tr_{B}(\Ket{A}\sum_i\langle ii|)$, since $\tr_{B}(\Ket{A}\sum_i\langle ii|)=\tr_{B}(A\otimes\idol\sum_i|ii\rangle\sum_{i'}\langle i'i'|)=A$, where $\tr_B$ is partial trace for subsystem $B$. Note that $\varrho_{\mathrm{in}}$ from  $\Ket{\varrho_{\mathrm{in}}}$ may not be a valid quantum state (i.e., $\varrho_{\mathrm{in}}$ may not be a semidefinite matrix).

There are two sub-cases with $M^{-1}$ existing. \textit{Sub-Case (1a):}  $M^{-1} \Ket{\varrho_{\rm t}}$ corresponds to a valid quantum state $\varrho_{\mathrm{in}}$, where $\varrho_{\mathrm{in}}=\tr_{B}(\Ket{\varrho_{\mathrm{in}}}\sum_i\langle ii|)=\tr_{B}(M^{-1} \Ket{\varrho_{\rm t}}\sum_i\langle ii|)$, in this case there is only one solution for the input state $\varrho_{\mathrm{in}}$; \textit{Sub-Case (1b):} there is no valid quantum state $\varrho_{\mathrm{in}}$ such that $\Ket{\varrho_{\mathrm{in}}}=M^{-1} \Ket{\varrho_{\rm t}}$, i.e., $\tr_{B}(M^{-1} \Ket{\varrho_{\rm t}}\sum_i\langle ii|)$ is not a valid quantum state, and thus the expected input state $\varrho_{\mathrm{in}}$ does not exist. All we need to do is that from $M$ we calculate its inverse matrix $M^{-1}$ and check whether $\delta:=\tr_{B}(M^{-1} \Ket{\varrho_{\rm t}}\sum_i\langle ii|)$ is a valid quantum state or not (if yes $\varrho_{\mathrm{in}}=\tr_{B}(M^{-1} \Ket{\varrho_{\rm t}}\sum_i\langle ii|)$, otherwise $\varrho_{\mathrm{in}}$ does not exist).

\begin{figure*}[htb]
\includegraphics[scale=0.65]{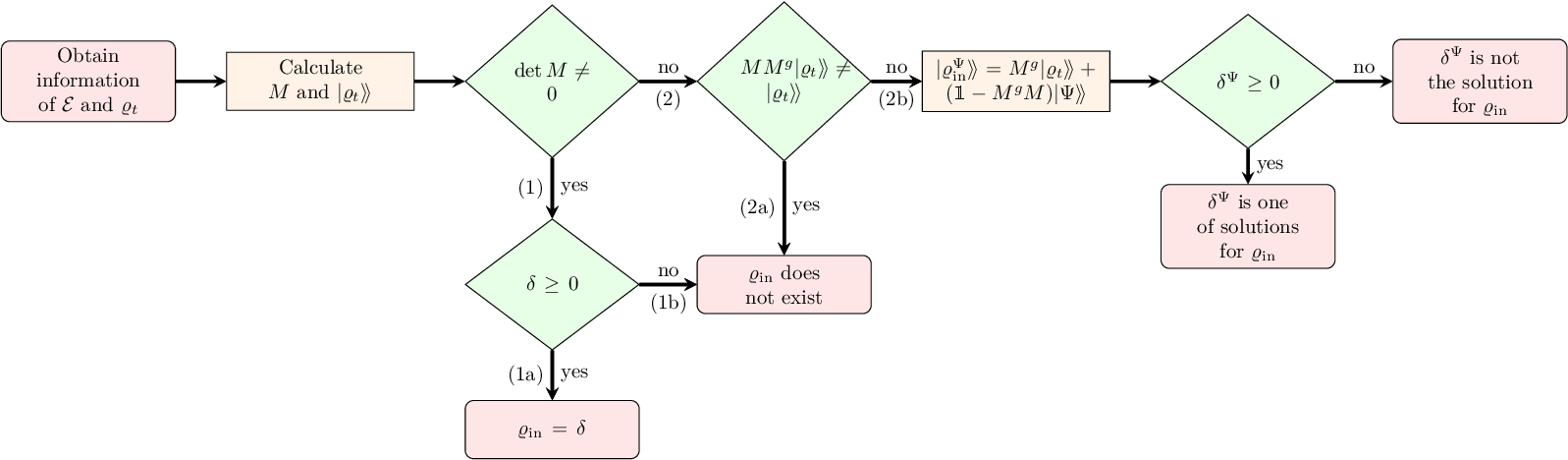}
\caption{The procedure for annlytically designing input state $\varrho_{\mathrm{in}}$ with a given quantum channel $\mathcal{E}$ and a target state $\varrho_{\rm t}$. If the whole system is just a single-partite system with a quantum channel, as in Eq. (\ref{one-party}), one can obtain $M:=\sum_i K_i\otimes K_i^*$, $\Ket{\varrho_{\rm t}}:=\varrho_{\rm t}\otimes \idol \sum_i|ii\rangle$,  $\delta:=\tr_{B}(M^{-1} \Ket{\varrho_{\rm t}}\sum_i\langle ii|)$, and $\delta^\Psi:=\tr_{B}(\Ket{\varrho_{\mathrm{in}}^\Psi}\sum_i\langle ii|)$. }\label{fig2}
\end{figure*}

\textit{Case (2).} The matrix $M:=\sum_i K_i\otimes K_i^*$ has no inverse matrix $M^{-1}$ (i.e., its determinant $\det M=0$).  There are two sub-cases as well.  \textit{Sub-Case (2a):} $M\Ket{\varrho_{\mathrm{in}}}=\Ket{\varrho_{\rm t}}$ has no solution for $\Ket{\varrho_{\mathrm{in}}}$ (i.e., $MM^g\Ket{\varrho_{\rm t}}\neq\Ket{\varrho_{\rm t}}$ \cite{james,inverse}, where $M^g$ is the Moore-Penrose pseudo-inverse  of $M$ \cite{inverse}), and thus in this sub-case  the input state $\varrho_{\mathrm{in}}$ does not exist.  Mathematically, the Moore-Penrose pseudo-inverse $A^g$ of a matrix $A$ is the most well known generalization of inverse matrix, which is unique for simultaneously satisfying the following four conditions, $AA^g A=A$, $A^g A A^g=A^g$, $(A A^g)^\dag=A A^g$, $(A^g A)^\dag=A^g A$, see \cite{Moore,Penrose}.  \textit{Sub-Case (2b):}  $M\Ket{\varrho_{\mathrm{in}}}=\Ket{\varrho_{\rm t}}$ has an infinite number of solutions for $\Ket{\varrho_{\mathrm{in}}}$ (i.e., $MM^g\Ket{\varrho_{\rm t}}=\Ket{\varrho_{\rm t}}$), and all the solutions can be written as $\Ket{\varrho_{\mathrm{in}}^\Psi}=M^g\Ket{\varrho_{\rm t}}+(\idol-M^g M)\Ket{\Psi}$, where $\Ket{\Psi}$ is an arbitrary vector with the same dimension as $\Ket{\varrho_{\rm t}}$ \cite{james,inverse}. For all the solutions of $\Ket{\varrho_{\mathrm{in}}^\Psi}$ one needs to check whether each $\delta^\Psi:=\tr_{B}(\Ket{\varrho_{\mathrm{in}}^\Psi}\sum_i\langle ii|)$ is a valid quantum state (if $\delta^\Psi\geq0$) or not ($\delta^\Psi$ has at least one negative eigenvalue).

In principle, for an arbitrary quantum channel $\mathcal{E}$ and target state $\varrho_{\rm t}$, we can always follow the above procedure by checking which case it belongs to, and analytically obtaining the expected input state $\varrho_{\mathrm{in}}$ if it exists. The above procedure has been shown in Fig. \ref{fig2}.

\textit{Example 1.} Let us consider one qubit system with the quantum channel being Pauli maps. Suppose Alice and Bob share a Pauli map $\mathcal{E}_p$, $\varrho_{\rm t}=\mathcal{E}_p(\varrho_{\mathrm{in}})=\sum_{i=0}^{3} p_i \sigma_i \varrho_{\mathrm{in}} \sigma_i^{\dag}$
where $\sigma_0$ is the Identity matrix, $\{\sigma_i\}_{i=1}^{3}$ are the Pauli matrices,  and $\sum_{i=0}^{3} p_i=1$, with $0\leq p_i\leq 1$.  Based on the definition of the matrix $M$, one can obtain $M=\sum_{i=0}^{3} p_i \sigma_i\otimes \sigma_i^*$.

\textit{Case (1).} The matrix $M=\sum_{i=0}^{3}p_i \sigma_i\otimes \sigma_i^*$ has an inverse matrix $M^{-1}$ (its determinant $\det M\neq0$), i.e. the following three conditions must hold simultaneously, (i) $q_1:=p_0+p_1-p_2-p_3\neq0$, (ii) $q_2:=p_0-p_1+p_2-p_3\neq0$, (iii) $q_3:=p_0-p_1-p_2+p_3\neq0$.
Suppose that the target output state is $\varrho_{\rm t}=\frac{1}{2}(\idol+\sum_{i=1}^{3}r_i \sigma_i)$ where $r_i=\tr(\sigma_i \varrho_{\rm t})$; from $\varrho_{\mathrm{in}}=\tr_{B}(M^{-1} \Ket{\varrho_{\rm t}}\sum_i\langle ii|)$ one has
\begin{equation}\label{rin}
    \varrho_{\mathrm{in}}=\frac{1}{2}\Big(\idol+\sum_{i=1}^{3}R_i\sigma_i\Big),
\end{equation}
where $R_i:=r_i/q_i$.
Clearly, $\varrho_{\mathrm{in}}$ in Eq. (\ref{rin}) is a valid quantum state if and only if $\sum_{i=1}^{3}R_i^2\leq1$, i.e., $(R_1,R_2,R_3)$ is a true Bloch vector.

\textit{Case (2).} The matrix $M=\sum_{i=0}^{3} p_i \sigma_i\otimes \sigma_i^*$ has no inverse matrix $M^{-1}$ (its determinant $\det M=0$), which means that at least one of  $\{q_i\}_{i=1}^{3}$ must be zero. We denote $k,l,m\in\{1,2,3\}$, and $k,l,m$ are different from each other.

(i) Only $q_k=0$ ($q_l$ and $q_m$ are not zero), from  $MM^g\Ket{\varrho_{\rm t}}=\Ket{\varrho_{\rm t}}$ one has $r_k=0$ for  the target output state $\varrho_{\rm t}=\frac{1}{2}(\idol+\sum_{i=1}^{3}r_i \sigma_i)$, and all the solutions of $\Ket{\varrho_{\mathrm{in}}}$ can be written as $\Ket{\varrho_{\mathrm{in}}^\Psi}=M^g\Ket{\varrho_{\rm t}}+(\idol-M^g M)\Ket{\Psi}$, where $\Ket{\Psi}$ is an arbitrary vector with the same dimension as $\Ket{\varrho_{\rm t}}$.  Thus, $\delta^\Psi=\tr_B(\Ket{\varrho_{\mathrm{in}}^\Psi}\sum_i\langle ii|)=\frac{1}{2}(\idol+\sum_{i=1}^{3}\tilde{R}_i \sigma_i)$,
where $\tilde{R}_l=r_l/q_l$, $\tilde{R}_m=r_m/q_m$, but $\tilde{R}_k$ can be an arbitrary real number. Furthermore, one can see that $\delta^\Psi\geq0$ if and only if $\sum_{i=1}^3 \tilde{R}_i^2\leq1$.

(ii) $q_k=q_l=0$ but $q_m\neq0$, from  $MM^g\Ket{\varrho_{\rm t}}=\Ket{\varrho_{\rm t}}$ one has $r_k=r_l=0$, and $\delta^\Psi=\frac{1}{2}(\idol+\sum_{i=1}^{3}R_{i}^{'} \sigma_i)$, with $R_m^{'}=r_m/q_m$, $R_k^{'}$ and $R_l^{'}$ can be arbitrary real numbers. Furthermore, one can see that $\delta^\Psi\geq0$ if and only if $\sum_{i=1}^3 {R_i^{'}}^2\leq1$.

(iii) If $q_1=q_2=q_3=0$, from  $MM^g\Ket{\varrho_{\rm t}}=\Ket{\varrho_{\rm t}}$ one has $r_1=r_2=r_3=0$ and $\delta^\Psi=\frac{1}{2}(\idol+\sum_{i=1}^{3}\tilde{R}_{i}^{'} \sigma_i)$,
with $\tilde{R}_1^{'}$, $\tilde{R}_2^{'}$ and $\tilde{R}_3^{'}$ arbitrary real numbers satisfying $\sum_{i=1}^3 ({\tilde{R}_i}^{'})^2\leq1$.

\subsection{Bipartite systems}
We have designed the input state if  Alice would like to send a special target state $\varrho_{\rm t}$ to Bob via a quantum channel. The whole system we considered is just a single-partite system.  Let us now assume that Alice and Bob would like to share an entangled target state $\varrho_{\rm t}^{AB}$, and this entangled state is initially prepared by Alice. So Alice needs to send one subsystem to Bob, and keep the other one. In this case, what initial state $\varrho_{\mathrm{in}}^{AB}$ should Alice prepare?

Suppose that there is a quantum channel between Alice and Bob. The quantum channel can be viewed as a CPTP map $\mathcal{E}$,  with the output state corresponding to an input state $\varrho$ being written in a Kraus form Eq. (1). Alice would like to share a special target state $\varrho_{\rm t}^{AB}$ with Bob. She can try to prepare an initial quantum state $\varrho_{\mathrm{in}}^{AB}$, and sends the subsystem B to Bob, such that
\begin{equation}\label{sbiCP}
\varrho_{\rm t}^{AB}=\idol\otimes\mathcal{E}(\varrho_{\mathrm{in}}^{AB})=\sum_i \idol\otimes K_i\varrho_{\mathrm{in}}^{AB} \idol\otimes K_i^\dag,
\end{equation}
where $K_i$ are operators satisfying the completeness relation $\sum_i K_i^\dag K_i=\idol$. Similarly, we use the notation
\begin{equation}\label{}
    \Ket{H^{AB}}:=H^{AB}\otimes\idol^{A'B'}\sum_{ij}|ijij\rangle^{ABA'B'},
\end{equation}
which relates the vector $\Ket{H^{AB}}$ and the operator $H^{AB}$.
Therefore, Eq. (\ref{sbiCP}) is equivalent to
\begin{eqnarray}\label{sbivec}
    \Ket{\varrho_{\rm t}^{AB}}&=&\Ket{\sum_i \idol\otimes K_i \varrho_{\mathrm{in}}^{AB} \idol\otimes  K_i^{\dag}}\nonumber\\
    &=&\sum_i \idol\otimes K_i\otimes \idol\otimes K_i^*\Ket{\varrho_{\mathrm{in}}^{AB}},
\end{eqnarray}
the last equation holds due to the definition of $\Ket{H^{AB}}$. Therefore, there are several cases for the choice of Alice's input state depending on the target output state $\varrho_{\rm t}^{AB}$ and the matrix
\begin{equation}\label{}
    M:=\sum_i \idol\otimes K_i\otimes \idol\otimes K_i^*.
\end{equation}

\textit{Case (1).} The matrix $M:=\sum_i \idol\otimes K_i\otimes \idol\otimes K_i^*$ has an inverse matrix $M^{-1}$  (i.e., its determinant $\det M \neq0$). Since $M^{-1}$ exists, from Eq. (\ref{sbivec})  we have
\begin{equation}\label{}
    \Ket{\varrho_{\mathrm{in}}^{AB}}=M^{-1} \Ket{\varrho_{\rm t}^{AB}},
\end{equation}
and from $\Ket{\varrho_{\mathrm{in}}^{AB}}$ one can obtain $\varrho_{\mathrm{in}}^{AB}$ by using
\begin{equation}\label{}
   H^{AB}=\tr_{A'B'}(\Ket{H^{AB}}\sum_{ij}\langle ijij|),
\end{equation}
because of the following equations,
\begin{eqnarray}
&&\tr_{A'B'}(\Ket{H^{AB}}\sum_{ij}\langle ijij|)\nonumber\\
&=&\tr_{A'B'}(H^{AB}\otimes\idol^{A'B'}\sum_{ij}|ijij\rangle\sum_{i'j'}\langle i'j'i'j'|)\nonumber\\
&=&H^{AB}.
\end{eqnarray}
It is worth noticing that $\varrho_{\mathrm{in}}^{AB}$ from  $\Ket{\varrho_{\mathrm{in}}^{AB}}$ may not be a valid quantum state.

There are two sub cases with $M^{-1}$ existing.

\textit{Sub-Case (1a):}  $M^{-1} \Ket{\varrho_{\rm t}^{AB}}$ corresponds to a valid quantum state $\varrho_{\mathrm{in}}^{AB}$, where
\begin{eqnarray}
\varrho_{\mathrm{in}}^{AB}&=&\tr_{A'B'}(\Ket{\varrho_{\mathrm{in}}^{AB}}\sum_{ij}\langle ijij|)\nonumber\\
&=&\tr_{A'B'}(M^{-1} \Ket{\varrho_{\rm t}^{AB}}\sum_{ij}\langle ijij|),
\end{eqnarray}
in this case there is only one solution for the input state $\varrho_{\mathrm{in}}^{AB}$;

\textit{Sub-Case (1b):} there is no valid quantum state $\varrho_{\mathrm{in}}^{AB}$ such that $\Ket{\varrho_{\mathrm{in}}^{AB}}=M^{-1} \Ket{\varrho_{\rm t}^{AB}}$, i.e., $\tr_{A'B'}(M^{-1} \Ket{\varrho_{\rm t}^{AB}}\sum_{ij}\langle ijij|)$ is not a valid quantum state, and thus the expected input state $\varrho_{\mathrm{in}}^{AB}$ does not exist. All we need to do now is that from $M$ we calculate its inverse matrix $M^{-1}$ and check whether
\begin{eqnarray}
\delta^{AB}:=\tr_{A'B'}(M^{-1} \Ket{\varrho_{\rm t}^{AB}}\sum_{ij}\langle ijij|)
\end{eqnarray}
is a valid quantum state or not (if yes $\varrho_{\mathrm{in}}^{AB}=\tr_{A'B'}(M^{-1} \Ket{\varrho_{\rm t}^{AB}}\sum_i\langle ijij|)$, otherwise $\varrho_{\mathrm{in}}^{AB}$ does not exist).

\textit{Case (2).} The matrix $M:=\sum_i \idol\otimes K_i\otimes \idol\otimes K_i^*$ has no inverse matrix $M^{-1}$ (i.e., its determinant $\det M=0$).  There are two sub cases as well.

\textit{Sub-Case (2a):} $M\Ket{\varrho_{\mathrm{in}}^{AB}}=\Ket{\varrho_{\rm t}^{AB}}$ has no solution for $\Ket{\varrho_{\mathrm{in}}^{AB}}$ (i.e., $MM^g\Ket{\varrho_{\rm t}^{AB}}\neq\Ket{\varrho_{\rm t}^{AB}}$ \cite{james}, where $M^g$ is the Moore-Penrose pseudo inverse  of $M$), and thus in this sub-case  the input state $\varrho_{\mathrm{in}}^{AB}$ does not exist;

\textit{Sub-Case (2b):}  $M\Ket{\varrho_{\mathrm{in}}^{AB}}=\Ket{\varrho_{\rm t}^{AB}}$ has infinite solutions for $\Ket{\varrho_{\mathrm{in}}^{AB}}$ (i.e., $MM^g\Ket{\varrho_{\rm t}^{AB}}=\Ket{\varrho_{\rm t}^{AB}}$), and all the solutions can be written as
\begin{equation}\label{}
\Ket{\varrho_{\mathrm{in}}^\Psi}=M^g\Ket{\varrho_{\rm t}^{AB}}+(\idol-M^g M)\Ket{\Psi},
\end{equation}
where $\Ket{\Psi}$ is an arbitrary vector with the same dimension as $\Ket{\varrho_{\rm t}^{AB}}$ \cite{james}. For all the solutions of $\Ket{\varrho_{\mathrm{in}}^\Psi}$ one needs to check whether each
\begin{equation}\label{}
    \delta^\Psi:=\tr_{A'B'}(\Ket{\varrho_{\mathrm{in}}^\Psi}\sum_{ij}\langle ijij|)
\end{equation}
is a valid quantum state or not.

In principle, for arbitrary quantum channels and target output states $\varrho_{\rm t}^{AB}$ we can always follow the above procedure by checking which case it belongs to, and analytically obtaining the expected input state $\varrho_{\mathrm{in}}^{AB}$ if it exists,  similar to the procedure shown in Fig. \ref{fig2}.

\textit{Example 2.} Let us consider a two-\textit{qutrit} system with only subsystem B passing through an amplitude damping channel. Assume that Alice and Bob share an amplitude damping channel $\mathcal{E}$,
\begin{equation}\label{}
    \varrho_{\rm t}^{AB}=\idol\otimes\mathcal{E}(\varrho_{\mathrm{in}}^{AB})=\sum_{i=0}^{2} \idol\otimes A_i \varrho_{\mathrm{in}}^{AB} \idol\otimes A_i^{\dag},
\end{equation}
where
\begin{eqnarray}
  A_0&=&|0\rangle\langle0|+\sqrt{1-\gamma}|1\rangle\langle1|+(1-\gamma)|2\rangle\langle2|,\\
  A_1&=&\sqrt{\gamma}|0\rangle\langle1|+\sqrt{2\gamma(1-\gamma)}|1\rangle\langle2|,\\
  A_2&=&\gamma|0\rangle\langle2|,
\end{eqnarray}
with $0\leq\gamma\leq1$. Assume that our target output state is
\begin{equation}\label{}
    \varrho_{\rm t}^{AB}=p|\psi^+\rangle\langle\psi^+|+(1-p)\frac{\idol}{9},
\end{equation}
where $|\psi^+\rangle=(|00\rangle+|11\rangle)/\sqrt{2}$, $\idol$ is the $9\times 9$ identity matrix, and $0\leq p \leq1$. Based on the definition of matrix $M$, one can obtain that
\begin{equation}\label{}
    M=\sum_i  \idol\otimes A_i\otimes \idol\otimes A_i^*.
\end{equation}

\textit{Case (1).} The matrix $M=\sum_i  \idol\otimes A_i\otimes \idol\otimes A_i^*$ has an inverse matrix $M^{-1}$ (i.e., its determinant $\det M\neq0$), which means $\gamma\neq1$.
From $\varrho_{\mathrm{in}}^{AB}=\tr_{A'B'}(M^{-1} \Ket{\varrho_{\rm t}^{AB}}\sum_{ij}\langle ijij|)$ one has
\begin{eqnarray}\label{rinAB}
    \varrho_{\mathrm{in}}^{AB}={\frac{1}{c}}\left(
\begin{array}{ccccccccc}
a_1 & 0 & 0 & 0 & b & 0 & 0 & 0 & 0 \\
0 & a_2 & 0 & 0 & 0 & 0 & 0 & 0 & 0 \\
0 & 0 & a_3 & 0 & 0 & 0 & 0 & 0 & 0 \\
0 & 0 & 0 & a_4 & 0 & 0 & 0 & 0 & 0 \\
b & 0 & 0 & 0 & a_5 & 0 & 0 & 0 & 0 \\
0 & 0 & 0 & 0 & 0 & a_6 & 0 & 0 & 0 \\
0 & 0 & 0 & 0 & 0 & 0 & a_7 & 0 & 0 \\
0 & 0 & 0 & 0 & 0 & 0 & 0 & a_8 & 0 \\
0 & 0 & 0 & 0 & 0 & 0 & 0 & 0 & a_9
\end{array}
\right),
\end{eqnarray}
where $a_1=2-6 \bar{\gamma} \gamma+p (7-12\gamma+3 \gamma^2)$, $a_2=a_8=2 \bar{p} (1 - 3 \gamma)$, $a_3=a_6=a_9=2\bar{p}$, $a_4=2 - 6\bar{\gamma} \gamma - p (2 + 3 \bar{\gamma} \gamma)$, $a_5=2 + 7 p - 3 (2 + p) \gamma$, $a_7=2 \bar{p} (1 - 3\bar{\gamma}\gamma)$, $b=9 p \bar{\gamma}^{3/2}$, $c=18 \bar{\gamma}^2$, $\bar{p}=1-p$ and $\bar{\gamma}=1-\gamma$.
It is easy to check that $\varrho_{\mathrm{in}}^{AB}$ in Eq. (\ref{rinAB}) is a valid quantum state if and only if the following two conditions hold simultaneously,
\begin{eqnarray}\label{}
   &&0\leq\gamma\leq\frac{1}{3}, \\
   &&0\leq p \leq\frac{2-6\bar{\gamma}\gamma}{2+3\bar{\gamma}\gamma}.
\end{eqnarray}

\textit{Case (2).} The matrix $M=\sum_i  \idol\otimes A_i\otimes \idol\otimes A_i^*$ has no inverse matrix $M^{-1}$ (i.e., its determinant $\det M=0$), which means $\gamma=1$. In this case, one can see that $MM^g\Ket{\varrho_{\rm t}^{AB}}\neq\Ket{\varrho_{\rm t}^{AB}}$ holds. Therefore, there is no solution for $\varrho_{\mathrm{in}}^{AB}$ when $M^{-1}$ does not exist.

\section{Numerical calculation by using SDP}
In the above section, we have provided an analytical result for designing input states with a given quantum channel and a target output state. Now we reconsider this problem by using the SDP numerical method. Assume that Alice and Bob share a quantum channel $\mathcal{E}$ described by Eq.~(\ref{CP}), and Alice and Bob obtain all the information of this quantum channel in advance. If Alice would like to send a special target state $\varrho_{\rm t}$ to Bob, to get the input state, we assume that there exists an input state $\varrho_{\mathrm{in}}$ such that Eq.~(\ref{one-party}) holds.

Let us choose operator-basis sets $\{F_k\}$ in the Hilbert-Schmidt spaces of Hermitian operators \cite{oS,oS2}, where $k=1,\cdots,d^2$, and $d$ is the dimension of the Hilbert space of $\varrho_{\rm t}$. These  basis sets $\{F_k\}$ satisfy $\tr(F_k F_{k'})=\delta_{kk'}$ and $\sigma=\sum_{k=1}^{d^2}\tr(\sigma F_k)F_k$, with $\sigma$ being an arbitrary $d\times d$ Hermitian matrix. For simplicity, we can choose $F_1=\idol/\sqrt{d}$. Therefore, Eq. (\ref{one-party}) is equivalent to $\tr[F_k\mathcal{E}(\varrho_{\mathrm{in}})]=\tr(F_k\varrho_{\rm t})$, with $k=1,\cdots,d^2$. Furthermore, one has $\tr[F_k\mathcal{E}(\varrho_{\mathrm{in}})]=\tr[\mathcal{E}^*(F_k)\varrho_{\mathrm{in}}]$, where $\mathcal{E}^*$ is a dual map of $\mathcal{E}$, and $\mathcal{E}^*(F_k)=\sum_i K_i^{\dag} F_k K_i$. Thus, Eq.~(\ref{one-party}) is equivalent to
\begin{equation}\label{sdp1}
    \tr[\mathcal{E}^*(F_k)\varrho_{\mathrm{in}}]=\tr(F_k\varrho_{\rm t}), \ \ k=1,\cdots,d^2.
\end{equation}
When $k=1$, Eq.~(\ref{sdp1}) is equivalent to the trace normalization condition of $\varrho_{\mathrm{in}}$,
\begin{equation}\label{sdp2}
    \tr\varrho_{\mathrm{in}}=1, \ \ \  \varrho_{\mathrm{in}}\geq0,
\end{equation}
Equations (\ref{sdp1}) and (\ref{sdp2}) form a natural SDP problem:
\begin{eqnarray}
 \mathrm{minimize} &&\  \tr(CX)\nonumber\\
 \mathrm{such \ that} &&\ \tr(B_kX)=b_k, \  \    \  k=1,\cdots,d^2\nonumber\\
                     &&\  X\geq0,\label{S1}
\end{eqnarray}
where $C=0$, $B_k=\mathcal{E}^*(F_k)$, and $b_k=\tr(F_k\varrho_{\rm t})$ for $k=1,\cdots,d^2$, $X=\varrho_{\mathrm{in}}$. Note that $C=0$ here. So the optimal value (always 0) does not depend on the choice of $X$ as long as it exists. This kind of SDP problem is called ``feasibility  problem", only to determine whether a feasible solution exists. The SDP problem (\ref{S1}) can be solved  by using the parser YALMIP \cite{yalmip} with the solvers, SEDUMI \cite{sedumi} or SDPT3 \cite{sdpt3,sdpt32}.

If there does not exist any input state $\varrho_{\mathrm{in}}$ such that  $\mathcal{E}(\varrho_{\mathrm{in}})=\varrho_{\rm t}$, one can still maximize the fidelity $F[\varrho_{\rm t},\mathcal{E}(\varrho_{\mathrm{in}})]$ between the target state $\varrho_{\rm t}$ and $\mathcal{E}(\varrho_{\mathrm{in}})$ over all possible input states $\varrho_{\mathrm{in}}$, where  the fidelity $F(\varrho_1,\varrho_2):=\tr[(\sqrt{\varrho_1}\varrho_2 \sqrt{\varrho_1})^{\frac{1}{2}}]=\|\sqrt{\varrho_1}\sqrt{\varrho_2}\|_1=\max_{U}|\tr(U\sqrt{\varrho_1}\sqrt{\varrho_2})|$ \cite{Rev1}, with $U$ being an arbitrary unitary operator and $\|\cdot\|_1$ being the trace norm. In particular, when the target state is a pure state $|\psi_t\rangle$, one has $F[|\psi_t\rangle, \mathcal{E}(\varrho_{\mathrm{in}})]=\sqrt{\langle\psi_t|\mathcal{E}(\varrho_{\mathrm{in}})|\psi_t\rangle}=\sqrt{\tr[\mathcal{E}^*(|\psi_t\rangle)\varrho_{\mathrm{in}}]}$. Therefore,
\begin{eqnarray}
\max_{\{\varrho_{\mathrm{in}}\}}F\big(|\psi_t\rangle, \mathcal{E}(\varrho_{\mathrm{in}})\big)=\sqrt{\lambda_{\mathrm{max}}},
\end{eqnarray}
where $\lambda_{\mathrm{max}}$ is the largest eigenvalue of $\mathcal{E}^*(|\psi_t\rangle)$ and $\varrho_{\mathrm{in}}$ is the corresponding eigenstate.

When the target state is a mixed state $\varrho_{\rm t}$, one can numerically calculate the maximum fidelity $F[\varrho_{\rm t},\mathcal{E}(\varrho_{\mathrm{in}})]$ via the SDP as \cite{SDP2,SDP22}
\begin{eqnarray}
 \mathrm{maximize} &&\  \frac{1}{2}\tr(P)+\frac{1}{2}\tr(P^\dag)\nonumber\\
 \mathrm{such \ that} &&\ \left(
\begin{array}{cc}
 \varrho_{\rm t} & P \\
 P^\dag & \mathcal{E}(\varrho_{\mathrm{in}})
\end{array}
\right)\geq0,\label{S2}
\end{eqnarray}
since the optimal value $\frac{1}{2}\tr(P)+\frac{1}{2}\tr(P^\dag)$ is equal to the fidelity $F[\varrho_{\rm t},\mathcal{E}(\varrho_{\mathrm{in}})]$.
One can use the parser YALMIP \cite{yalmip} with the solvers SEDUMI \cite{sedumi} and PENBMI \cite{penbmi}, to solve the SDP problem (\ref{S2}).

Now we reconsider the Pauli map $\mathcal{E}_p$ in Example 1 with $p_0=0.7$ and $p_1=p_2=p_3=0.1$ in the Appendix B. We have numerically generated 10,000 random target states $\varrho_{\rm t}$. Using the above SDP, we found that there are 75.16\% target states which can be perfectly error pre-compensated (in this case $F[\varrho_{\rm t},\mathcal{E}(\varrho_{\mathrm{in}})]=1$ and our analytical results coincide with SDP results) ,  89.3\% target states with the fidelity $F[\varrho_{\rm t},\mathcal{E}(\varrho_{\mathrm{in}})]>0.99$, and 100\% target states with the fidelity $F[\varrho_{\rm t},\mathcal{E}(\varrho_{\mathrm{in}})]>0.90$.


\section{Advantages and shortcomings of QEPC}
The advantage of the QEPC method is that Bob does not need to do anything after the quantum process tomography of a given quantum channel. If Alice would like to send a target state to Bob, she can design an error pre-compensated input state according to Fig. \ref{fig2}, and Bob would just receive the output state without any \textit{priori} information of the target state. As mentioned before, in the QECC and QER methods, Bob needs to do correcting or recovery operations, which more or less depend on a \textit{priori}  knowledge of the target state.

Let us now compare the QEPC scheme with the QECC method.
Suppose we encode a single qubit information in an $n$-qubit quantum code which can correct arbitrary errors on any single qubit, with the total error probability $p$. Using the $n$-qubit quantum code, the fidelity satisfies (see Section 10.3.2 in \cite{Rev1})
\begin{equation}
F=\sqrt{(1-p)^{n-1}(1-p+np)}=1-\frac{\binom{n}{2}}{2}p^2+O(p^3).\label{F}
\end{equation}
Thus, when $n$ is large, the total probability of all errors $p$ should be sufficiently small. Otherwise, the $n$-qubit quantum code cannot improve the fidelity of the state protected by the code. We present the following example to show the case. 


\begin{figure*}[htb]
\subfigure{
\includegraphics[scale=1.25]{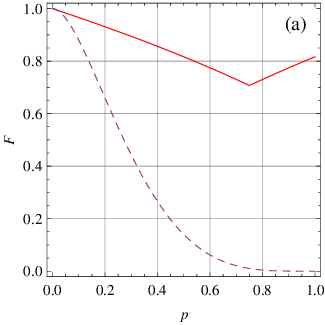}}
\subfigure{
\includegraphics[scale=1.27]{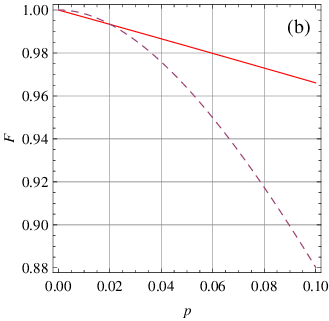}}
\caption{\textbf{(a)} Comparison of the fidelity $F_d'$ using the QEPC scheme and the fidelity $F_d$ using the Shor code. The red line denotes the fidelity $F_d'$ using the QEPC scheme, and the dashed line denotes the fidelity $F_d$ using the Shor code.  \textbf{(b)} Details of  figure (a) when $0.1\geq p\geq 0$.
}\label{fig4}
\end{figure*}

\textit{Example 3.} Let us consider the depolarizing channel, $\mathcal{E}_d(\varrho_{\mathrm{in}})=(1-p)\varrho_{\mathrm{in}}+p/3(\sum_{i=1}^3\sigma_i\varrho_{\mathrm{in}}\sigma_i)$.
If the target state is $|0\rangle$, using the Shor code $|0_L\rangle=(|000\rangle+|111\rangle)(|000\rangle+|111\rangle)(|000\rangle+|111\rangle)/(2\sqrt{2})$,
we can calculate the fidelity based on Eq. (\ref{F}) with $n=9$,
\begin{equation}\label{}
  F_d=\sqrt{(1-p)^8(1+8p)},
\end{equation}
one can obtain the details in the Appendix C.
Let us now design an input state $\varrho_{\mathrm{in}}=\frac{1}{2}(\idol+\sum_{i=1}^{3}R_i\sigma_i)$ and maximize the fidelity
\begin{equation}\label{}
F_d'=\max_{\{\varrho_{\mathrm{in}}\}}\sqrt{\langle0|\mathcal{E}_b(\varrho_{\mathrm{in}})|0\rangle}=\sqrt{1/2+|1/2-2p/3|}.
\end{equation}
When $1\geq p> 0.0204$,  $F_d'=\sqrt{1/2+|1/2-2p/3|}>\sqrt{(1-p)^8(1+8p)}=F_d$. See Fig. \ref{fig4} for details.

Furthermore, one may use $[[5,1,3]]$ code instead of the Shor code. In this case, $n=5$ and  the fidelity based on Eq. (\ref{F}) is
\begin{equation}\label{}
  F_d''=\sqrt{(1-p)^4(1+4p)}.
\end{equation}
One can find that when $1\geq p >0.0782$, $F_d'=\sqrt{1/2+|1/2-2p/3|}>\sqrt{(1-p)^4(1+4p)}=F_d''$.

However, the QEPC method has its shortcomings. First of all, the QEPC scheme needs the full information of quantum channels by quantum process tomography, but QECC methods do not need it. Moreover, when the target states are pure states, we can maximize the fidelity between the output mixed state and the target pure state; but, in general, the fidelity is less than one because there is no measurement or recovery operation in the QEPC scheme. Another limitation is that the QEPC scheme is not resistant under small deviations from the calculated channel noise and the actual channel effects. For instance, if the channel is strongly time-dependent or there are no exact methods to obtain the Kraus operators, the QEPC is not suitable.

\section{Discussions and conclusions}
In Fig.~\ref{fig1}, the initial state $\varrho_{\mathrm{in}}$ of the QEPC model, if it exists, can be an arbitrary pure state or a mixed state. Will the difficulty of the initial state preparation balance off the benefit brought by getting rid of error recovery? Actually, it depends on the physical realization and the scheme to be realized. Consider this special case: if Bob has no ability to do any operation to the output state, then Alice's pre-compensation is better than Bob's recovery procedure. On the other hand, even in the standard encoding-error-recovery model, Alice needs to do initial state preparation and encoding as well.

Compared with the active protecting methods in Refs. \cite{Knill,Knill2,Viola}, our QEPC scheme is also applied before error events occurred. The difference is that, the input state $\varrho_{\mathrm{in}}$ is usually the target state $\varrho_{\mathrm{t}}$ in the active protecting methods in Refs. \cite{Knill,Knill2,Viola}, however, in the QEPC model $\varrho_{\mathrm{in}}$ is not $\varrho_{\mathrm{t}}$ in general.

Let us compare the analytical and the numerical methods. First, following Fig. \ref{fig2}, one can always analytically find solutions of $\varrho_{\mathrm{in}}$ if these exist. Furthermore, if there exist more than one solution of $\varrho_{\mathrm{in}}$, all solutions of $\varrho_{\mathrm{in}}$ can be analytically obtained. But the SDP numerical methods will only find one solution of $\varrho_{\mathrm{in}}$. Second, the analytical procedure and the SDP (\ref{S1}) are designed for perfect error
pre-compensation. Nevertheless, the SDP (\ref{S2}) is designed to find the maximum fidelity, which is not a perfect error
pre-compensation when the maximum fidelity is not one. Third, if there is no solution for $\varrho_{\mathrm{in}}$, the analytical procedure and the SDP (\ref{S1}) will get nothing. But, using the SDP (\ref{S2}) one can always find the maximum fidelity between the target state $\varrho_{\rm t}$ and $\mathcal{E}(\varrho_{\mathrm{in}})$, although the maximum fidelity is less than 1.

A practical scenario for the QEPC method is polarization-encoding quantum key distribution via optical fibers. In Refs. \cite{oe1,oe2,oe3,ol}, the authors experimentally tested and compensated the polarization random drifts, which usually compensate the drifts only for the states $\{|H\rangle, |V\rangle, |45\rangle,|-45\rangle\}$  after the quantum channel of optical fibers. Here we introduce the QEPC method for pre-compensation of the errors before the quantum channels. One may use the QEPC model to pre compensate the polarization random drifts in experiments of  quantum key distribution via optical fibers. 

In conclusion, we have proposed QEPC method for quantum noisy channels. The required input state can be analytically and numerically obtained if it exists. If the required input state does not exist, we can find the input state, such that the output state is as close as possible to the target output state by SDP. In this work, there is no encoding or decoding operation, and we do not combine the QEPC model with other strategies, such as dynamical decoupling \cite{Rev4,DD1,DD2,DD3,DD4,DD5,DD6}. For future research, one may use encoding and decoding (or even recovery) operations and  dynamical decoupling in the QEPC model.

\section*{ACKNOWLEDGMENTS}
We thank anonymous referees for  useful suggestions,  and thank  Simon Devitt, Otfried G\"uhne, Daniel Herr, Adam Miranowicz, Franco Nori, and Jiang Zhang for helpful discussions and comments. C.Z. is funded by the National Natural Science Foundation of China (Grant No. 11734015),  
and K.C. Wong Magna Fund in Ningbo University. H.Y. is supported by the Research Grants Council of Hong Kong (RGC, Hong Kong) (Grant No. 538213). R.D. was involved in this work when he worked at University of Technology Sydney.

\setcounter{equation}{0}
\renewcommand{\theequation}{A\arabic{equation}}

\section*{APPENDIX A: CALCULATION OF $\Ket{\varrho_{\rm t}}$}
We use the notation
\begin{equation}\label{}
    \Ket{A}:=A\otimes\idol\sum_i|ii\rangle=\idol\otimes A^T\sum_i|ii\rangle=\sum_{ij}A_{ij}|ij\rangle,
\end{equation}
with $A=\sum_{ij}A_{ij}|i\rangle\langle j|$ \cite{horn,D'Ariano}. $A^T$ denotes transposition of $A$, and $\idol$ is the identity operator. We suppose that there exists an input state $\varrho_{\mathrm{in}}$ such that
\begin{equation}\label{sone-party}
    \varrho_{\rm t}=\mathcal{E}(\varrho_{\mathrm{in}})=\sum_i K_i \varrho_{\mathrm{in}} K_i^{\dag},
\end{equation}
which is equivalent to \cite{horn,D'Ariano}
\begin{equation}\label{svec}
    \Ket{\varrho_{\rm t}}=\Ket{\sum_i K_i \varrho_{\mathrm{in}} K_i^{\dag}}=\sum_i K_i\otimes K_i^*\Ket{\varrho_{\mathrm{in}}}.
\end{equation}
To obtain the last equation, we use the definition of $\Ket{A}$,
\begin{eqnarray}
 \Ket{\varrho_{\rm t}}&=&\Ket{\sum_i K_i \varrho_{\mathrm{in}} K_i^{\dag}}\nonumber\\
 &=&\sum_i K_i \varrho_{\mathrm{in}} K_i^{\dag}\otimes\idol \sum_j|jj\rangle\nonumber\\
 &=&\sum_i K_i \varrho_{\mathrm{in}}\otimes  K_i^{*}\sum_j|jj\rangle\nonumber\\
 &=&\bigg(\sum_i K_i\otimes K_i^{*}\bigg) \bigg(\varrho_{\mathrm{in}}\otimes\idol\bigg)\sum_j|jj\rangle\nonumber\\
 &=&\sum_i K_i\otimes K_i^*\Ket{\varrho_{\mathrm{in}}},
\end{eqnarray}
where the third equation holds since $A\otimes\idol\sum_j|jj\rangle=\idol\otimes A^T\sum_j|jj\rangle$.

\setcounter{equation}{0}
\renewcommand{\theequation}{B\arabic{equation}}

\section*{APPENDIX B: EXAMPLE USING SEMIDEFINITE PROGRAMS}
Let us reconsider Example 1 in the main text using the semidefinite program (9). Let us assume that Alice and Bob share a Pauli map $\mathcal{E}_p$,
\begin{equation}\label{}
    \varrho_{\rm t}=\mathcal{E}_p(\varrho_{\mathrm{in}})=\sum_{i=0}^{3} p_i \sigma_i \varrho_{\mathrm{in}} \sigma_i^{\dag},
\end{equation}
where $\sigma_0$ is the Identity matrix, $\{\sigma_i\}_{i=1}^{3}$ are Pauli matrices, $\sum_{i=0}^{3} p_i=1$, with $0\leq p_i\leq 1$.  For simplicity, we can choose $F_1=\frac{\idol}{\sqrt{2}}$, $F_2=\frac{\sigma_1}{\sqrt{2}}$, $F_3=\frac{\sigma_2}{\sqrt{2}}$, $F_4=\frac{\sigma_3}{\sqrt{2}}$. Using $B_k=\mathcal{E}^*(F_k)$, one can obtain
\begin{eqnarray}
B_1&=&\mathcal{E}_p^*(F_1)=\frac{\idol}{\sqrt{2}},\nonumber\\
B_2&=&\mathcal{E}_p^*(F_2)=\frac{\sigma_1}{\sqrt{2}}(p_0+p_1-p_2-p_3)=\frac{\sigma_1}{\sqrt{2}}q_1,\nonumber\\
B_3&=&\mathcal{E}_p^*(F_3)=\frac{\sigma_2}{\sqrt{2}}(p_0-p_1+p_2-p_3)=\frac{\sigma_2}{\sqrt{2}}q_2,\nonumber\\
B_4&=&\mathcal{E}_p^*(F_4)=\frac{\sigma_3}{\sqrt{2}}(p_0-p_1-p_2+p_3)=\frac{\sigma_3}{\sqrt{2}}q_3,\nonumber
\end{eqnarray}
i.e.,
\begin{equation}\label{}
    B_i=\mathcal{E}_p^*(F_i)=\frac{\sigma_i}{\sqrt{2}}q_i,  \ \ \ \ (i=0,1,2,3)
\end{equation}
where
\begin{eqnarray}
&&q_0:=p_0+p_1+p_2+p_3=1,\\
&&q_1:=p_0+p_1-p_2-p_3,\\
&&q_2:=p_0-p_1+p_2-p_3,\\
&&q_3:=p_0-p_1-p_2+p_3.
\end{eqnarray}
Suppose that the target output state is
\begin{eqnarray}
\varrho_{\rm t}=\frac{1}{2}(\idol+r_1 \sigma_x+r_2\sigma_y+r_3\sigma_z).
\end{eqnarray}
From $b_k=\tr(F_k\varrho_{\rm t})$ one has
\begin{eqnarray}
b_1=\tr(F_1\varrho_{\rm t})=\frac{1}{\sqrt{2}},\\
b_2=\tr(F_2\varrho_{\rm t})=\frac{r_1}{\sqrt{2}},\\
b_3=\tr(F_3\varrho_{\rm t})=\frac{r_2}{\sqrt{2}},\\
b_4=\tr(F_4\varrho_{\rm t})=\frac{r_3}{\sqrt{2}}.
\end{eqnarray}
Therefore,  the conditions of the SDP problem (9) in the main text $\tr(B_k\varrho_{\mathrm{in}})=b_k$ become
\begin{eqnarray}
\tr(\varrho_{\mathrm{in}})&=&1,\\
q_1\tr(\sigma_1\varrho_{\mathrm{in}})&=&r_1,\\
q_2\tr(\sigma_2\varrho_{\mathrm{in}})&=&r_2,\\
q_3\tr(\sigma_3\varrho_{\mathrm{in}})&=&r_3.
\end{eqnarray}
When $q_i\neq0$ simultaneously, this SDP problem becomes the Case (1) of Example 1 (which used the analytical method) in the main text. When at least one  $q_i=0$, this SDP problem becomes Case (2) of Example 1. In Case (2), if there exist more than one solution of $\varrho_{\mathrm{in}}$, all solutions of $\varrho_{\mathrm{in}}$ can be analytically obtained, but this SDP numerical method will only find one solution of $\varrho_{\mathrm{in}}$. The MATLAB code for the semidefinite program (11) is simple.
%
%
%
%
%
%
%
%
%
%
%
One can use the parser YALMIP \cite{yalmip} with the solvers, SEDUMI \cite{sedumi} or SDPT3 \cite{sdpt3,sdpt32}. The numerical results coincide with the analytical results.

Furthermore, let us reconsider Example 1 in the main text using the semidefinite program (11). The MATLAB code for the semidefinite program (11) is simple.
%
%
%
%
%
%
%
%
%
%
We have used the parser YALMIP \cite{yalmip} with the solvers, SEDUMI \cite{sedumi} and PENBMI \cite{penbmi}, where PENBMI is useful as designed for solving optimization
problems (as ours) with bilinear matrix inequality constraints.

The numerical results coincide with the analytical results and the numerical results from the semidefinite program (11). For instance,  for the Pauli map with $p_0=0.7$ and $p_1=p_2=p_3=0.1$, we have numerically generated 10,000 random target states $\varrho_{\rm t}$. Using the above MATLAB code, we found that there are 75.16\% target states which can be perfectly  error pre-compensated (in this case $F[\varrho_{\rm t},\mathcal{E}(\varrho_{\mathrm{in}})]=1$ and our analytical results coincide with SDP results) ,  89.3\% target states with the fidelity $F[\varrho_{\rm t},\mathcal{E}(\varrho_{\mathrm{in}})]>0.99$, and 100\% target states with the fidelity $F[\varrho_{\rm t},\mathcal{E}(\varrho_{\mathrm{in}})]>0.90$.


\setcounter{equation}{0}
\renewcommand{\theequation}{C\arabic{equation}}

\section*{APPENDIX C: FIDELITIES OF THE QEPC SCHEME AND QUANTUM ERROR-CORRECTING CODES}

Let us now consider the depolarizing channel, which is  a special case of Pauli maps,
\begin{equation}\label{}
    \mathcal{E}_d(\varrho_{\mathrm{in}})=(1-p)\varrho_{\mathrm{in}}+\frac{p}{3}(\sigma_1\varrho_{\mathrm{in}}\sigma_1+\sigma_2\varrho_{\mathrm{in}}\sigma_2+\sigma_3\varrho_{\mathrm{in}}\sigma_3).
\end{equation}
If the target state is $|0\rangle$, we use the Shor code
\begin{equation}
 |0_L\rangle=\frac{(|000\rangle+|111\rangle)(|000\rangle+|111\rangle)(|000\rangle+|111\rangle)}{2\sqrt{2}}.
\end{equation}
Suppose the depolarizing channel with parameter $p$ acts independently on each of the qubits, giving rise to a joint action on all 9 qubits of the Shor code, then
the quantum state after both the noise and error-correction is (see Section 10.3.2 in \cite{Rev1})
\begin{equation}\label{}
    \varrho_{\mathrm{QECC}}=[(1-p)^9+9p(1-p)^8]|0_L\rangle\langle0_L|+\cdots.
\end{equation}
Therefore, we can calculate the fidelity (see Section 10.3.2 in \cite{Rev1}),
\begin{eqnarray}\label{}
   F_d&=&\sqrt{\langle0|\varrho_{\mathrm{QECC}}|0\rangle}\nonumber\\
   &=&\sqrt{(1-p)^8(1+8p)}.
\end{eqnarray}
On the other hand, let us design an input state
\begin{equation}\label{Srin}
    \varrho_{\mathrm{in}}=\frac{1}{2}\Big(\idol+\sum_{i=1}^{3}R_i\sigma_i\Big),
\end{equation}
and maximize the fidelity
\begin{eqnarray}
 F_d'
 &=&\max_{\{\varrho_{\mathrm{in}}\}}\sqrt{\langle0|\mathcal{E}_b(\varrho_{\mathrm{in}})|0\rangle}\nonumber\\
 &=&\max_{\{\varrho_{\mathrm{in}}\}}\sqrt{(1-p)\langle0|\varrho_{\mathrm{in}}|0\rangle+\frac{2p}{3}\langle1|\varrho_{\mathrm{in}}|1\rangle+\frac{p}{3}\langle0|\varrho_{\mathrm{in}}|0\rangle}\nonumber\\
 &=&\max_{R_3}\sqrt{(1-p)\frac{1+R_3}{2}+\frac{p}{3}(1-R_3)+\frac{p}{3}\frac{1+R_3}{2}}\nonumber\\
 &=&\max_{R_3}\sqrt{\frac{1}{2}+R_3\Big(\frac{1}{2}-\frac{2p}{3}\Big)}\nonumber\\
 &=&\sqrt{\frac{1}{2}+\Big|\frac{1}{2}-\frac{2p}{3}\Big|}.
\end{eqnarray}
In Fig. 3 in the main text, we show that when $1\geq p>0.0204$, 
\begin{equation}
 F_d'=\sqrt{\frac{1}{2}+\Big|\frac{1}{2}-\frac{2p}{3}\Big|}>\sqrt{(1-p)^8(1+8p)}=F_d.
\end{equation}
The Shor code can improve the fidelity only when $p$ is extremely small ($0<p<0.0204$).

Similarly, if we use $[[5,1,3]]$ code instead of the Shor code. In this case, $n=5$ and  the fidelity is
\begin{equation}\label{}
  F_d''=\sqrt{(1-p)^4(1+4p)}.
\end{equation}
One can find that when $1\geq p >0.0782$,  
\begin{equation}
 F_d'=\sqrt{\frac{1}{2}+\Big|\frac{1}{2}-\frac{2p}{3}\Big|}>\sqrt{(1-p)^4(1+4p)}=F_d''.
\end{equation}

\end{document}